# Galaxy Group Analysis - A Robust Discriminator Between Cosmological Models: Cold + Hot Dark Matter and CDM Confront CfA1

Richard Nolthenius
UCO/Lick Observatory, University of California, Santa Cruz, CA 95064

Anatoly A. Klypin
Department of Astronomy, New Mexico State University, Las Cruces, NM 88001

Joel R. Primack
Physics Department, University of California, Santa Cruz, CA 95064

**Abstract**

We present techniques for identifying and analyzing galaxy groups and show that they provide a powerful and robust discriminator between cosmological models. We apply these methods to high-resolution particle-mesh (PM) N-body simulations of structure formation in three $\Omega = 1$ cosmological models; Cold plus Hot Dark Matter (CHDM) with $\Omega_{\rm cold} = 0.6$, $\Omega_\nu = 0.3$, and $\Omega_{\rm baryon} = 0.1$ at b=1.5 (COBE normalization) and two CDM models; b=1.5 and b=1.0 (COBE normalization). Groups are identified with the adaptive friends-of-friends algorithm of Nolthenius (1993). Our most important conclusions are: (1) Properties of groups are a powerful and robust discriminator between these Gaussian cosmological models whose spatial and velocity dispersion properties differ on ∼Mpc scales. We test robustness against several methods for assigning luminosity to dark matter halos, for merging CfA1 data, and for breaking up massive dark matter halos to correct for the overmerger problem. (2) When allowance is made for the higher than typical large scale power present in the CfA1 data, CHDM at our $\Omega_\nu = 0.30$ produces slightly too many groups and too high a fraction of galaxies in groups, while the fraction grouped in CDM is far too low. A slightly lower $\Omega_\nu$ would appear to produce excellent agreement with all measures, save one: for all simulations, median group sizes are up to a factor of 1.7 larger than equivalently selected CfA1 groups. This is either a real difference or perhaps due to residual resolution limitations. (3) The standard group M/L method applied to our $\Omega = 1$ simulations gives $\Omega \simeq 0.1$ for CHDM and CfA1, and $\Omega \simeq 0.35$ for CDM. We show quantitatively how three different effects conspire to produce this large discrepancy, and conclude that low observed Ω's need not argue for a low Ω universe. When overmergers are broken up, the median virial-to-DM mass $M_{vir}/M_{DM}$ of 3D selected groups is $\sim 1$ for all simulations. Groups with $M_{DM} > 10^{14}\ M_\odot$ appear virialized in all simulations. We measure global velocity biases $b_v$ similar to previous studies. Within 3D-selected groups, CHDM and CDM b=1.5 show a stronger bias of $b_v = 0.7 - 0.8$, while CDM b=1.0 shows group $b_v$'s $\simeq 1$.

*Subject headings:* cosmology: theory — dark matter — large scale structure of the universe — galaxies: formation — galaxies: clustering

## 1. Introduction

The simplest viable scenario of cosmic evolution begins with an $\Omega = 1$ expansion seeded from inflation with Gaussian primordial fluctuations with a Harrison-Zel'dovich scale invariant spectrum. These fluctuations later collapse via gravitational instability to form structure from galaxies on up. This picture has been remarkably resilient in the face of steadily mounting observations (Gorski, *et al.* 1994, Dekel 1994). The nature of the dark matter will govern how structure forms, and is thus in principle recoverable from the statistics of such structure (and hopefully, of course, from direct detection as well). Evidence of structure on very large scales in the late '80's began a series of efforts which have all but ruled out standard Cold Dark Matter (CDM, e.g. Baugh & Efstathiou 1993). This has motivated the investigation of alternative models with the desirable properties of significant power on large scales and small power on small scales, including Cold + Hot Dark Matter (CHDM). As recently emphasized by Pogosyan (1994), standard CDM's demise leaves CHDM as the only remaining theory compatible with the simplest and most aesthetic versions of inflation. Even more interesting, recent neutrino oscillation experiments provide preliminary evidence that at least one neutrino species indeed has a cosmologically important mass (Caldwell 1994, Primack *et al.* 1994).

Beginning with Klypin, *et al.* (1993, hereafter KHPR), we explore in this series of papers the consequences of a universe dominated by Cold + Hot Dark Matter and compare it with standard Cold Dark Matter and with observations. The CHDM model with $\Omega_\nu = 0.3$ has already shown good agreement with observations of the galaxy correlation function (Baugh & Efstathiou 1993), galaxy pairwise velocities (Somerville *et al.* 1995) and bulk velocities (KHPR), the cluster-cluster correlation function (Holtzman & Primack 1993; Klypin & Rhee 1994), the variance and skewness of the Abell/ACO cluster distribution (Plionis, *et al.* 1994), the amplitude of the power spectrum from POTENT reconstruction of the local density field (Seljak & Bertschinger 1994), the QDOT-IRAS Redshift Survey power spectrum (Feldman, Kaiser, & Peacock 1994, Fisher *et al.* 1993), quietness of the local Hubble flow (Schlegel, *et al.* 1994), the x-ray properties of clusters vs. redshift (Bryan, *et al.* 1994) and an initial analysis of the properties of galaxy groups (Nolthenius, Klypin & Primack 1994 a; hereafter NKP94). More controversial, (Ghigna, *et al.* 1994) find the standard CDM void probability function is in good agreement with observations in the Perseus-Pisces Supercluster region while that for CHDM is too high. deLapparent, Geller & Huchra (1986), however, find standard CDM produces voids too small, relative to those in the CfA1 Survey (Davis, *et al.* 1982), suggesting CHDM would be in better agreement here. Also, while CHDM agrees better with the APM galaxy angular correlation function $\omega(\theta)$ than does standard or tilted CDM, CHDM's $\omega(\theta > 3^\circ)$ is still slightly too low (Yepes, *et al.* 1994). Finally, the Hubble constant $H_0$ can be no larger than $\sim 50$ km s$^{-1}$Mpc$^{-1}$ to avoid problems with overproducing clusters and, as for all $\Omega = 1$ models, staying within cosmic age constraints. Therefore, if the current observational evidence for $H_0 \simeq 80$ km s$^{-1}$Mpc$^{-1}$ is confirmed, then CHDM in its simplest and most aesthetic form is ruled out. Throughout this paper, we assume $H_0 = 50$ km s$^{-1}$Mpc$^{-1}$. Aside from this, the easiest way to rule out CHDM is to find massive collapsed objects at high redshift, as CHDM forms structure significantly later than most competing models. These observations are currently underway, and indeed the first results are pushing the preferred $\Omega_\nu$ down to perhaps 0.2 (Klypin, *et al.* 1994a).

In this paper, we show how the statistics of galaxy groups provide a powerful discriminator between cosmological models. We identify and analyze groups from the 100 Mpc $512^3$ particle-mesh simulations described in Klypin, Nolthenius, & Primack 1994b (hereafter KNP). The box size was chosen to give a good statistical sample of galaxy groups. With a mesh size of 195 Kpc (perhaps typical of an $\sim L^*$ galaxy dark matter halo), we do not have the resolution to study individual galaxies. However, typical galaxy groups have virial radii of $\sim 1-2$ Mpc and begin to be resolved. In NKP94 we presented our methods and an initial analysis of groups in the CHDM and CDM simulations, and in the CfA1 Survey. This paper provides a more complete analysis, and includes corrections for effects not considered earlier.

It can be validly argued that the properties of real groups depend sensitively on the properties of real galaxies, since galaxies are still the tracers for which we have the best observational statistics, and so the problem of realistically identifying galaxies is not so easy to circumvent. For example, group virial radii depend strongly on the spatial distribution of individual galaxies. Until more realistic large-scale simulations become feasible, the only way to deal with this problem is to add in by hand properties which are well determined physically but poorly understood or difficult to simulate (e.g. luminosity), and to do so equitably across all competing cosmological models. We have addressed the resolution problem by filtering the observational data to the same resolution as the simulations. Fortunately, tests on most of our chosen statistics show little sensitivity to the relatively poor spatial resolution and related uncertainties, as we show in the present paper. A more significant problem is that of "overmerging" (Katz & White 1993, Gelb & Bertschinger 1993). Dark matter (DM) halos are extended, soft objects which merge easily. Galaxies undergo dissipative collapse to denser, smaller-cross section objects which merge at perhaps only half the DM merger rate (Evrard, Summers, & Davis 1993, hereafter ESD). We considered several schemes for breaking up our most massive halos. While with reasonable assumptions group rms velocities and fractions grouped appear insensitive to the details of halo breakup, we will show it is nevertheless possible to find breakup prescriptions that do lead to significantly different group properties.

We analyze groups selected from two different versions of our catalogs. In order to gain insight into the properties of our models using all available information, we select groups from the complete 100 Mpc simulation boxes using full three dimensional (3D) information ("box groups"). To make meaningful, direct comparisons with the CfA1 data, we first make magnitude limited "sky catalogs" in redshift space from each simulation, and then identify groups using identical criteria.

The construction of the $512^3$ grid of cells within the 100 Mpc box and the simulation calculation methods are described in KNP. In NKP94, we presented a robust and discriminatory statistic, rms group velocity $v_{\rm gr}$ (the rms velocity of all galaxies within the group, i.e. the conventional "velocity disperion") vs. the fraction of catalog galaxies in groups $f_{\rm gr}$ , and compared CHDM with CDM and with the CfA1 Survey data. Here, we present a more complete description of the construction of improved galaxy redshift catalogs, including corrections for some effects not considered earlier, analyze the properties of 3D-selected groups from the full box, and present other comparisons between these catalogs and observations. An accompanying computer visualization video sequence (Brodbeck, *et al.* 1994) compares visually the differences between the simulations, sky catalogs, and the real universe. CDM structures are rather puffy filaments, with clusters at the intersections. By contrast, CHDM is seen to have remarkably delicate filaments which are reminiscent, in a picturesqe way, of a well prepared egg flower soup. Blurring in redshift space reduces the visual impression of these differences, yet their effect remains powerful on the statistics studied here.

Grouping in redshift space is done using the adaptive algorithm of Nolthenius (1993; N93), although tests with the original algorithm of Nolthenius and White (1987; NW) gave essentially the same results but with slightly lower group velocity dispersions. Recently, Frederic (1995a,b) has claimed that the NW grouping algorithm seriously underestimates group velocity dispersions and hence M/L ratios when compared with simulation groups. However, these studies fail to emphasize the importance of the underlying cosmology to setting the grouping algorithm's appropriate link in redshift. Using only a low bias standard CDM simulation, as Frederic did, will indeed show that the NW redshift link normalization of $V_5 = 350$ km s$^{-1}$ is too low, as already shown in NKP94, due to the high pairwise velocities in this model. The fault lies not in the algorithm, but in the assumption that standard CDM is an appropriate calibrating cosmology for determining the merits of grouping algorithms used on real observations. Likewise, discrepancies he finds with the Ramella, *et al.* (1989; RGH) CfA2 group results vanish when one realizes the CfA2 dataset should not be assumed to have CDM-like properties.

## 2. Construction of the Galaxy Halo Catalogs

A detailed description of the simulation calculations is given in NKP94 and KNP. Briefly, we use a particle mesh code on a $512^3$ grid, with $256^3$ cold and $2 \times 256^3$ hot particles. The cold particle mass is $2.9 \times 10^9\ M_\odot$ and $4.1 \times 10^9\ M_\odot$ for CHDM and CDM, respectively. As before, we refer to the CDM b=1 (COBE normalization, see Smoot, *et al.* 1992) and b=1.5 simulations as CDM1 and CDM1.5, respectively. The two CHDM simulations, CHDM1 and CHDM2, are both at b=1.5 (COBE normalization), and differ only in their initial conditions. The $\mathrm{CHDM}_1$, CDM1, and CDM1.5 simulations all began with the same random number set describing the amplitudes of the initial waves perturbing the particles. It was later found that this random number set had abnormally high power on large scales. The power spectrum was a factor of $\sim 2$ higher than typical on scales comparable to the box size. The probability of this occuring in any given realization was estimated at $\sim 10\%$. A second CHDM simulation, $\mathrm{CHDM}_2$, began with a much more typical spectrum. This fact will be important later in interpreting the comparisons with CfA1 data. As it turns out, there is good evidence that CfA1 has unusual large scale power as well. The value of beginning each competing model with the same random perturbation set is that it guarantees the same large scale structures will emerge in each, so that differences between simulations will solely be due to differences in the underlying physics of the evolution and not to cosmic variance.

A galaxy halo (boldly referred to hereafter as a "galaxy") is defined as a mesh cell with a sufficiently high dark particle mass overdensity $(\delta\rho/\rho)_{cut}$ at the end of the simulation (z=0). We quantify the mass of a halo by the mass contained within a cell (or sometimes 3x3x3 cell) boundary. This is obviously crude. The boundaries are arbitrary and no attempt is made (nor is it possible) to include only the gravitationally bound particles; our spatial and force resolution is too poor to justify such refinements. Nevertheless, as long as galaxy luminosity monotonically rises with 1-cell mass, our results are insensitive to how mass is quantified. When our purpose is to optimally delineate structure we keep all cells above $(\delta\rho/\rho)_{cut} = 30$, giving N=29151 halos in CHDM1, N=37,164 in CDM1, N=45,592 in CDM1.5, and N=29,795 in $\mathrm{CHDM}_2$. When constructing magnitude limited redshift catalogs and attempting to match CfA1 galaxy number densities, experiments showed that $(\delta\rho/\rho)_{cut} = 80 - 150$ was best, cutting galaxy totals by about 70%. Note that this is slightly less than the $(\delta\rho/\rho)_{cut} = 170$ corresponding to virialization, (e.g. Kaiser 1986), and seems reasonable if one assumes virialization should actually apply to a denser core of material closer to the visible galaxy. Since each cell is 1/512 of the 100 Mpc box, or 195 Kpc , a cell would most properly correspond to a dark matter halo surrounding a typical $L^*$ galaxy. Two measures of the mass of such cells were calculated; the dark particle mass contained within the cell, and the dark particle mass contained within a $3^3$ cell cube (3-cell) centered on the cell of interest.

### 3. Breaking Up Massive Halos

One uncertainty in our earlier results is that the overmerging of dark matter galaxy halos may have significantly lowered the rms velocities of galaxies within groups, and perhaps also lowered the total fraction of galaxies in groups. The dominant galaxy will tend to sit near the group center and have low center-of-mass velocity. By contrast, baryonic dissipation causes earlier collapse into several smaller but higher velocity galaxies (Katz & White 1993) whose merging rate is slower (ESD). If, as has been argued by some, the overmerging seen in numerical simulations is mostly due to poor mass resolution, our results should be relatively secure. Our mass resolution is quite good; an $L^*$ galaxy has about 100 cold particles and twice as many hot particles. Assuming the more likely probability that the lack of dissipation is indeed the dominant source of overmerging, we now ask above what halo mass $M_{bu}$ does it become important? Katz & White (1993) find that some individual halos could be as massive as $1.5\times10^{14}\ M_\odot$ . However, using reasonable M/L ratios, cooling efficiencies, and star formation rates, they argue most halos above $M_{bu} = 5\times10^{13}\ M_\odot$ should be broken up, and that $M_{bu}$ could be as low as $2.8\times10^{12}\ M_\odot$ . Gelb (1992) finds too many halos above $V_{circ} = (\frac{GM}{r})^{1/2} = 350$ km s$^{-1}$ which, for $r$ corresponding to our cell size, is equivalent to $3.4\times10^{12}\ M_\odot$ . ESD also address this issue, and with better mass resolution they find a lower $M_{bu}$ limit of $7\times10^{11}\ M_\odot$ . Our most massive CHDM halo has 3-cell mass of $5\times10^{13}\ M_\odot$ (or a 1-cell mass of $5\times10^{12}\ M_\odot$ ). To date, ESD is the best published work available on galaxy formation in groups and clusters in a dark matter background. Their Figure 12a provides a rough relation between dark matter halo mass and the number of galaxy-like objects ("globs" in their nomenclature) within the halo in standard CDM. While this figure applies to z=1, their most recent results show the number of globs stays fairly constant to z=0, at least in standard CDM (Evrard 1994). They define the halo's mass as the mass within the sphere containing an average overdensity corresponding to virialization: $\delta\rho/\rho = 170$. Assuming unevolving halo size from z=1 to the present, the cosmological expansion factor of 2 then gives a corresponding $\delta\rho/\rho = 2^3 \times 170 = 1360$ for $M_{bu}$ in our z=0 simulations. We found the radius $r_{eff}$ and mass $M_{eff}$ of this sphere by determining the radii of spheres encompassing the volume of our 1-cell and 3-cell masses, and assuming density fell linearly from the 1-cell sphere radius to the 3-cell sphere radius. The maximum radius allowed to enclose a single breakup candidate correponded to the sphere enclosing a $2^3$-cell volume, or halfway to the nearest allowable halo. This was done to avoid double counting some of the exterior mass. We then broke up halos whose mass $M_{eff}$ within a sphere of average overdensity $\delta\rho/\rho = 1360$ was greater than $7 \times 10^{11}\ M_\odot$ . The result of this procedure was $N_{om}$ = 138 ($CHDM_1$), 157 ($CHDM_2$), 571 (CDM1.5), and 658 (CDM1) halos identified as overmergers and suitable for breaking up into fragments. If mass is roughly proportional to light, observed galaxies suggest that fragments should be assigned Schechter-distributed masses. Dissipational hydrodynamic codes also produce gaseous galaxy-like objects with Schechter distributed masses (Evrard 1994). We therefore constrained the fragments to follow a Schechter distribution with the characteristic mass $M^* = \langle M/L \rangle L^* = 4.3 \times 10^{11}\ M_\odot$ (CHDM) or $1.08 \times 10^{12}\ M_\odot$ (CDM). $L^*$ is $4.3 \times 10^{10}\ L_\odot$ from the CfA1 and $\langle M/L \rangle$ is the median M/L of the simulation halos; 10 for CHDM and 25 for CDM. The faint end slope was set to the merged CfA1's $\alpha = -1.26$ (see §6). Each overmerger was replaced with fragments of total mass M, where M was the virialized overdensity mass $M_{eff}$ calculated above, minus the mass expected to be in fragments below the overdensity limit $(\delta\rho/\rho)_{cut}$ selected for the simulation (and thus too faint to see); 80 for both CHDM and 156 for both CDM simulations (see §6). For all simulations, $16 - 19\%$ of the integrated Schechter function is contained in fragments below $(\delta\rho/\rho)_{cut}$. For each overmerger, the brightest fragment was given a mass $M_{bf} = (M_{bu} M_{eff})^{1/2}$ (Evrard 1994) and placed at the original overmerger's position. Remaining masses were then randomly selected from a Schechter distribution and randomly placed into any halo which could accept it without overfilling. This continued until the next random fragment mass was too large to be added to any halo. At this point, all halos were $\sim 99\%$ full of fragments. Each set of fragments for a given halo was then ordered by mass and then sequentially placed as close as possible to the parent halo cell while still enforcing the 2-cell closest neighbor resolution limit. Thus the most massive fragments were placed closest to the original DM center. Each fragment was then given a randomly oriented, random Gaussian velocity with dispersion equal to the rms velocity $V^{om}_{neigh}$ of all halos within 1 Mpc of the overmerged halo (or, in the few cases this did not include at least 4 halos, out to the 4-th nearest halo). The median values $\langle V^{om}_{neigh} \rangle_{med}$ of these neighborhood rms velocities were $\sim 300$ km s$^{-1}$ for all models, but individual overmergers ranged as high as $\sim 1200$ km s$^{-1}$ (CHDM) to $\sim 2000$ km s$^{-1}$ (CDM1). The most massive fragment was left at the original overmerger's position. Observations indicate that the brightest (assumed the most massive) galaxies in groups and poor clusters are moving at a center-of-mass velocity only $\sim 0.25$ that of their lower luminosity neighbors (Bird 1994). We therefore multiplied the brightest fragment's velocity by 0.25. The fragments of a given overmerger do not, at this point, satisfy conservation of momentum. In the frame of the parent overmerger, the net momentum of the N fragments with masses $m_i$ is $\sum_i^N m_i \vec{v_i} = \vec{p}$. To enforce momentum conservation we correct each fragment velocity by adding a velocity differential $\vec{\delta v} = -\vec{p} (\sum_i^N m_i)^{-1}$. Finally, the masses were rescaled by $M_{1cell}/M_{eff}$ so that they followed the 1-cell convention used for all other cataloged halos. 3-cell masses were found in a similar way.

The simulation halos we analyze below have all been subjected to this, our preferred breakup scheme. We also experimented with other breakup schemes. These are described in detail below, and summarized in Table 1.

The first alternative scheme, hereafter "Method 1" made use of ESD's Figure 12a, giving the number of galaxies per DM halo vs. DM halo mass. $M_{bu}$ was again $7 \times 10^{11}$ $M_{\odot}$ , but this time the halo mass was simply assumed to be the 1-cell mass; generally a bit lower than the mass inside a sphere of average overdensity 1360. We broke each overmerger candidate into ESD's nominal number of fragments, giving them essentially equal masses (when ordered by mass, each fragment was an arbitrary 10% more massive than the previous fragment). Velocity assignments were as before, except rather than using all neighbors inside a distance of 1 Mpc, we used the 10 nearest neighbors, whatever their distance (typically out to a distance of $\sim 2$ Mpc ; as large as a medium sized group). The most massive fragment's velocity was not reduced. Placements again enforced the 2-cell nearest neighbor limit, but this time without regard to putting the most massive fragment closest to the DM center. The largest overmerger in any simulation spawned only 7 fragments, and the large majority of overmergers produced only 2 or occasionally 3 fragments. By essentially maximizing the mass of each fragment, this prescription guaranteed (albeit unintentionally) the highest possible number of fragments surviving the magnitude limit and making it into the final sky catalogs (see §6). Since these galaxies survived in close pairs or groups, it also significantly raised the fraction of galaxies in groups.

The second scheme ("Method 2") made the more reasonable assumption of Schechter-distributed masses, randomly selected and assigned with the same $M^*$ parameters as our adopted scheme, but with ESD's "glob" faint end slope of $\alpha = -1.35$. $M_{bu}$ was the same as for our adopted scheme. We did not constrain the brightest fragment's mass or velocity (except by enforcing momentum conservation, as before). Fragments were positioned randomly, regardless of mass, but enforcing the 2-cell limit. By not insuring at least one reasonably massive fragment, the result was a larger number ($\sim 30$ on average) of low mass fragments, relatively fewer of which ultimately survived the magnitude limit for inclusion into the sky catalogs. Velocities were randomly sampled from a Gaussian distribution with dispersion equal to the circular velocity $V_c^{om} = (GM_{eff}/r_{eff})^{1/2}$ associated with the virialization mass and radius described in our final method above.

Our preferred scheme assigns fragments their nearest neighbors' rms velocity $V_{neigh}^{om}$ rather than the circular velocity just described, for several reasons. First, the most appropriate mass and radius to use are poorly defined. Second, galaxies clearly form early, and, in the dense environments common to these overmergers, it is the larger tidal field of the group which seems likely to eventually determine final fragment velocities. Finally, one of our goals was to evaluate how well the virial theorem measures mass in these systems, and enforcing dynamically determined velocities would bias these results. Figure 1 shows a scatter plot of $V_c^{om}$ vs. $V_{neigh}^{om}$ for each overmerger in the CHDM and CDM simulation boxes. The $V_c^{om}$ and $V_{neigh}^{om}$ distributions differ substantially. However, their median values are very similar $\langle V_c^{om} \rangle_{med} \simeq \langle V_{neigh}^{om} \rangle_{med}$, as shown in Table 2. Circular velocities are slightly higher than $V_{neigh}^{om}$; by $\sim 5\%$ for CHDM and by $\sim 15\%$ for CDM. Thus, using circular velocities would likely have raised our final group rms velocities by only a few percent, more so for CDM than for CHDM. (However, it might also be argued that a more appropriate dynamical velocity would be lower than $V_c$, closer to the virial velocity, which would be lower by $\sim 2^{-1/2}$). Note also that the extended tail of high $V_{neigh}^{om}$ is especially pronounced for the CDM models, and is related to CDM having higher pairwise velocities than CHDM (KHPR).

Another approach ("Method 3") is to leave the simulation halos alone and instead attempt to put real galaxies back into "overmerged" halos by merging them, i.e. taking a luminosity weighted averaged of their positions and redshifts, then combining luminosities. In this case, one wants to merge CfA1 galaxies which are within about $1.5 - 2$ cells ($\sim 350$ Kpc ) on the sky, and within a velocity separation corresponding to the virial velocity of these massive halos, i.e. about $200 - 300$ km s$^{-1}$. Since this is typical of the rms velocity of a medium sized group, Method 3 turns out to correspond closely with the merging scheme already done in NKP94 (although our motivation then was actually to correct for spatial resolution, not overmerging). As we will see, the NKP94 results are quite close to the results of our more careful analysis here. We regard Methods 2 and 3, and especially Method 1, as less realistic than our adopted procedure. The assumptions for each of our breakup methods are summarized in Table 1.

**Table 1. Summary of Overmerger Breakup Methods**

| Method | Assumptions |
|---|---|
| Preferred | $M_{bu} = 7 \times 10^{11}\ M_\odot$ , $M$ inside $\delta\rho/\rho > 1360$ |
| | Schechter fragment masses M; $M^* =< M/L >_{sim} L^*_{CfA}$, $\alpha = \alpha_{CfA} = -1.26$ |
| | $\vec{v}_{fragment}$ =Gaussian ($\sigma$,$\vec{v}^{\dagger}_{om}$), $\sigma$=rms of galaxies within 1 Mpc or nearest 4 |
| | Fragments positioned by mass; higher mass closer to center |
| | Brightest fragment mass $M_{bf} = (M_{bu} M_{eff})^{1/2}$, at overmerger center, |
| | and velocity $\vec{v}_{bf} = 0.25 \times \vec{v}_{fragment}$ |
| Method 1 | $M_{bu} = 7 \times 10^{11}\ M_\odot$ , $M$ inside $\delta\rho/\rho > 1360$ |
| | No. fragments from ESD Fig 12a; masses, if ordered, each 90% of previous mass |
| | $\vec{v}_{fragment}$ =Gaussian ($\sigma$,$\vec{v}^{\dagger}_{om}$), $\sigma$=rms of 10 nearest galaxies |
| | Fragment positions not mass-dependent |
| Method 2 | $M_{bu} = 7 \times 10^{11}\ M_\odot$ , $M$ inside $\delta\rho/\rho > 1360$ |
| | Schechter fragment masses M; $M^* =< M/L >_{sim} L^*_{CfA}$, $\alpha = \alpha_{ESD} = -1.35$ |
| | $\vec{v}_{fragment}$ =random Gaussian ($\sigma$,$\vec{v}^{\dagger}_{om}$), $\sigma = (GM_{eff}/reff)^{1/2}$ |
| | Fragment positions not mass-dependent |
| Method 3 | Leave simulation overmerges alone |
| | Merge CfA1 grouped galaxies within $r_{proj}$ = 235 Kpc , i.e. NKP94 results |

fragment positions enforced 2-cell limit, for all methods
† $\vec{v}_{om}$ = velocity of original overmerger

Figure 2 shows the 1-cell mass distribution of the halos. The no breakup halos follow a power law of slope $d\log N/d\log M = -1.33$. We stress that the 1-cell and 3-cell masses, being defined by arbitrary boundaries, are not particularly physical measures of dark matter halo mass, and in our analysis are used only as stepping stones to luminosity. All breakup methods lead to a fairly sharp cutoff in masses at the upper end. Note that Method 1 makes an especially noticable pile of fragment masses just below the breakup mass limit $M_{bu}$.

All of these breakup schemes may in fact overestimate the number of fragments. Other purely dissipationless simulations (e.g. Carlberg 1994) find that dense DM cores are suprisingly persistent within virialized clusters, suggesting that overmerging may be less significant than generally believed. Also, if $M^*/10$ is equivalent to a luminosity of $L^*/10$, ESD and Evrard (1994) find too many fragments by an order of magnitude, when compared to observations. Our preferred procedure produces even more fragments (a median of 8 fragments per overmerger for both CDM simulations and 14 for both CHDM simulations, with a maximum of $\sim 140$ for all). Our luminosity assignment method (see §6) prevents overpopulation of visible galaxies, but perhaps does not prevent too high a fraction of visible galaxies which are fragments. In any case, as we'll see, our favored statistics turn out to be insensitive to reasonable breakup assumptions, and quite similar to our no breakup results.

## 4. Properties of Box Groups

We first describe groups selected using full 3D spatial information from a complete sample of all galaxies above a rather low 1-cell mass cutoff. Our procedure for constructing redshift space "sky catalogs" that can be compared to CfA1 data are described in §6. A group is defined as the galaxies within a bounding surface of constant number overdensity. Our grouping scheme is the standard "friends-of-friends" algorithm. The critical link distance $D$ is found from the N galaxies in the box above the mass cut by $D = D_0(l^3/N)^{1/3}$, where $l$ is the length of the box and $D_0$ is the fraction of the mean interparticle spacing, a dimensionless parameter. For each galaxy, all neighbors within a distance D are linked. Each of the newly linked neighbors is in turn searched, until no more members are found. We adopted $D_0 = 0.36$ (corresponding to a number overdensity for selected groups of $\sim 20$), for our comparisons, yielding a fraction grouped of $\sim 70\%$ for CHDM and $\sim 60\%$ for CDM. This overdensity limit best corresponds with the redshift selected groups described here and in earlier work (e.g. Huchra & Geller 1982, Nolthenius & White 1987 (NW), RGH, Nolthenius 1993 (N93)). In constructing groups near the box boundaries, the periodic images of galaxies were also considered during linking. This avoided the problem of clipping groups near the boundaries artificially.

Table 2 shows properties of groups made from the full box. The CHDM $(\delta\rho/\rho)_{cut} = 80$ and CDM $(\delta\rho/\rho)_{cut} = 156$ boxes will later be assigned luminosities and generate observational sky catalogs, as these halo density cuts best match CfA1 galaxy densities (see §6). We refer to these below as our fiducial boxes. To facilitate comparison between CHDM and CDM at the same halo mass density cut, we also include $(\delta\rho/\rho)_{cut} = 80$ CDM box results below. Note that these box groups are made from a sample containing a high fraction of low mass galaxies and groups, unlike the later groups to be made from magnitude limited sky catalogs.

**Table 2. Properties of Box Groups**

| simulation | $CHDM_2$ | $CHDM_1$ | CDM1.5 | CDM1 |
|---|---|---|---|---|
| halo $(\delta\rho/\rho)_{cut}$ | 80 | 80 | 156 | 156 |
| $N^*_{halo}$ | 11305(8585) | 10898(8134) | 14636(8503) | 16050(7247) |
| $N^{**}_{om}$ | 157 | 138 | 571 | 658 |
| $V_c^{om}$ km s$^{-1}$ | 309 | 316 | 315 | 322 |
| $V_{neigh}^{om}$ km s$^{-1}$ | 283 | 319 | 266 | 291 |
| $N_{grps}$ | 640(620) | 575(575) | 736(680) | 617(514) |
| $\langle M_{vir}/M_{DM}\rangle_{med}$ | 1.26(1.02) | 1.43(1.18) | 1.27(1.40) | 1.44(1.96) |
| $f_{\rm gr}$ † | .71 | .73 | .58 | .64 |
| DM frac in groups‡ | .15(.23) | .14(.21) | .23(.38) | .27(.50) |

quantities in () are for no-breakup case
* number of halos in the box
** number of overmerged halos
† fraction of halos which are in groups
‡ fraction of total DM which is inside the mean harmonic radius $r_h$ of groups

For our fiducial boxes, the mean spacing between groups is $\sim 9-12$ Mpc for all simulations. However, groups are strongly concentrated along filaments surrounded by large voids (see Brodbeck, *et al.* 1994), so that the distance to the nearest neighboring group is generally smaller; 5 Mpc on average, and 4.7 Mpc median.

Figure 3 shows the average distribution of mass around box groups in all simulations. Figure 3a also shows separate densities of the hot and cold fractions for $CHDM_2$. Since virial radii $r_h$ ranged from $\sim 0.6$ to $\sim 8$ Mpc, we minimized smearing in distance by binning the mass about each group as a function of the non-dimensional radius $r/r_h$. The result is an unweighted average for all groups. The profile is approximately exponential $\rho \propto e^{-r/r_h}$. Beyond the core, a power law of slope $-2$ (i.e. an isothermal sphere) fits the CHDM curves reasonably well. Simulations of galaxy clusters in a range of cosmologies (Crone, Evrard and Richstone 1994) find $\Omega = 1$ cosmologies also show an isothermal DM distribution, while $\Omega \simeq 0.2$ cosmologies do not. $CHDM_2$ densities are $\sim 10-15\%$ higher than $CHDM_1$. CDM densities, shown in Figure 3b, are also exponential. Groups made of halos over $(\delta\rho/\rho)_{cut} = 156$ have densities on average $\sim 70\%$ higher than those made of halos over $(\delta\rho/\rho)_{cut} = 80$. The density distribution shows no detectable change of slope beyond the radius defined by the halos. Indeed, out at $2r_h$ the density is still an order of magnitude above the critical density. This mass is likely still strongly bound to the group. Figure 4 shows the cumulative mass for the fiducial boxes. Note that there is still a significant amount of mass from $r_h < r < 2r_h$ which appears to be part of the groups; $M(2r_h)$ is $2.7M(r_h)$ for CHDM and $2.0M(r_h)$ for CDM. CDM box groups appear to contain a larger fraction of their DM mass at small $r$. Figure 4b shows the cumulative mass density distribution around sky groups. Note that smearing due to grouping in redshift space all but erases the differences in the $M(r)$ trends between simulations. Sky groups are on average much brighter and more massive than box groups, due to the magnitude limit. These more massive groups have an even broader distribution of surrounding dark matter, and $M(2r_h)/M(r_h)$ is $2.8 - 3.0$ for all simulation groups.

In the Cold + Hot Dark Matter picture, the cold particles fall into gravitational potential wells first, later to be followed by the hot particles after they cool. This offset in time continues until the present and leads to a higher fraction of cold particles in the cores of the groups, as shown in Figure 5. Interestingly, the $CHDM_1$ simulation's higher power on large scales seems to lead to a higher contrast between cold and hot densities, perhaps through earlier collapse and higher concentration of cold particles today. Data inside 195 Kpc is unresolved and not plotted.

We also calculated the virial masses, DM mass content, and velocity bias for these groups. These will be discussed in connection with determinations of $\Omega$ in §8.

## 5. Merging the CfA1 Catalog

Several issues need to be considered before we can make realistic comparisons between simulations and CfA1 data. First is their differing spatial resolution. The galaxy identification scheme requires that a cell identified as a galaxy be at a local density maximum. Thus, the nearest possible neighboring halo will be two cells away. We require the CfA1 data to show the same resolution, on average, before we can reliably identify equivalent groups. We therefore attempt to merge CfA1 galaxies which are sufficiently close. In NKP94 we used a simple merging scheme which assumed an isotropic orientation for the separation vector between all galaxies within a group and merged all galaxy pairs separated on the sky by less than $2r_{cl}/\pi$, where $r_{cl}$ is the spacing between nearest possible neighbors averaged over the positions of all possible nearest neighbors, and $2/\pi$ projects this onto the sky. The isotropic assumption is actually incorrect. Isotropy would only be true if the depth of the group were as small as $r_{cl}$. Here we perform a more careful merging which more closely models the resolution of the simulations. As such merging may be important for future comparisons between simulations and observations, we describe it in some detail below.

Consider a simulation cell tagged to be a galaxy halo. The nearest possible neighboring halo will lie on a cube 5 cells on a side ("neighbor cube"), such that inside this cube the 26 cells immediately adjacent to the central halo in question cannot be halos. To get an estimate of the mean separation between nearest neighboring halos, one could simply average the lengths of the separation vectors between the parent halo and all cells on the neighbor cube. To improve this estimate, we first weight each cell on the neighbor cube by the frequency that it is actually occupied in the simulations. For example, we find the closest neighbor cells, i.e. those offset from the central cell in one coordinate only, are $\sim 36\%$ more likely to be occupied than those on the edges and corners (average over all simulations). The resulting mean 3D separation between closest neighbors is $r_{cl} = 0.509$ Mpc . Now consider two concentrations of dark matter within the simulations. If the separation $r$ between the centroids of these concentrations is $r < r_{cl}/2$, our halo finder will define these as a single, merged halo. If $r > r_{cl}/2$, they will be identified as two halos separated by $r_{cl}$. Thus, one should seek to merge CfA1 galaxies closer than $r_{cl}/2$ (those between $r_{cl}/2$ and $r_{cl}$ are balanced, on average, by those between $r_{cl}$ and $1.5 r_{cl}$).

The last complication is that for real galaxies only sky-projected separations are known. Clearly group members in the simulation may be arbitrarily close together on the sky, given the right projection. The probability that they are actually close in 3D depends on the depth of the group relative to the pair's separation on the sky. Since candidate mergers will have sky separations of order $r_{proj} \leq 0.25$ Mpc and a typical group's depth is of order $1.5 - 2$ Mpc , the probability of merging is actually low and the isotropic estimate of NKP94 results in too many CfA1 mergings. To model the probabilities properly, we constructed a suite of artificial groups of various sizes (parameterized by the mean pairwise separation of the member galaxies $r_p$), with randomly positioned galaxies. Their density followed roughly the density of halos in groups; an $r < 0.55$ Mpc isodensity core surrounded by a $\rho \propto r^{-2}$ envelope, truncated so the resulting group boundary had an aspect ratio of 1:1:2 with the long axis along the line of sight. This elongation corresponds to the observed median aspect ratio of simulation groups in redshift spacce at the fiducial link parameters. We used these simulated groups to calculate the probability $P_{merge}(r_p, r_{proj})$ that the pair should be merged by tallying those pairs close in 3D as well as in sky projection. The resulting probability curves vs. $r_{proj}$ could be well fit with a set of second order polynomials. The output of this exercise was then an interpolation table of polynomial coefficients for each of 10 $r_p$'s spanning the range of observed CfA1 group $r_p$'s. For a median sized group, two galaxies separated on the sky by $r < r_{cl}$ had a probability for merging of order 0.1. We then looked at each N93 CfA1 group and binary galaxy pair, rolled the dice, and either merged or left alone each pair with $r_{proj} < r_{cl}$. After several trials, we adopted one which was typical, resulting in 39 group members and 5 binaries being merged. Note that this is only 2% of the galaxies.

## 6. Constructing Sky Projected Redshift Catalogs

For each galaxy in the fiducial simulation boxes, radial velocities $v_r$ in direction $\hat{r}$ at distance $r$ were calculated from

$$v_r = (\vec{v} - \vec{v_0}) \cdot \hat{r} + H_0 r, \qquad (1)$$

where the the observer is placed on a "home galaxy" with peculiar velocity $\vec{v_0}$. For the $\mathrm{CHDM}_1$ and $\mathrm{CHDM}_2$ simulations, we chose 6 home galaxies satisfying the following conditions: (a) the local galaxy density in redshift space ($V < 750$ km s$^{-1}$) is within a factor of 1.5 of the merged CfA1 galaxy density (though still usually on the low side), and (b) the closest Virgo-sized cluster is 20 Mpc away in distance. The eight most massive clusters ("Virgo"'s) in the $\mathrm{CHDM}_1$ box were centered about, and within a factor of $2-3$ of, the adopted Virgo mass of $M_{virgo} = 2.5 \times 10^{14}\ M_{\odot}$ . Corresponding $\mathrm{CHDM}_2$ clusters were less massive. Since the CDM1 and CDM1.5 boxes had the same initial random wave amplitudes as $CHDM_1$, the locations of large clusters and filaments were essentially the same as for $CHDM_1$. In order to further remove noise due to cosmic variance when comparing the different cosmological models, we chose the 6 viewing locations within CDM1 and CDM1.5 to be on the halos nearest to the viewing halo coordinates found for $CHDM_1$. The resulting home galaxies had an average peculiar velocity of $V_{pec} \simeq 800$ km s$^{-1}$. We retained all data out to $V = 12000$ km s$^{-1}$. Since our 100 Mpc box corresponds to 5000 km s$^{-1}$ for $H_0 = 50$ km s$^{-1}$Mpc$^{-1}$, this required periodically replicating each simulation box on all of its faces. However, at greater distances only the brightest galaxies are retained in our magnitude limited catalogs, so the replica boxes are sampled sparsely.

To simulate boundary effects yet still retain a large sample of the sky, we excised a 20° wide "zone of avoidance", leaving a solid angle of $A = 10.384$ sr. For some comparisons, it is important to match the CfA1 Survey geometry, despite the much poorer statistics. We therefore generated catalogs which had the local "Virgo" rotated to its CfA1 right ascension and declination, and the CfA1 latitude and declination limits imposed, giving a solid angle of $A = 2.66$ sr. The 10.384 sr catalogs are hereafter referred to as the "sky catalogs", the 2.66 sr catalogs are referred to as the "CfA-sky catalogs", and the merged CfA1 catalog as simply "CfA1".

An unavoidable problem in comparing all numerical simulations of large scale structure with real observations is how to assign luminosities to masses. How nature does this is still understood only in outline, and it is not obvious how our 1-cell masses should be assigned optical luminosities. We have therefore made the simple but plausible assumption that the baryonic mass in each 195 Kpc $^3$ cell turns into stars in such a way that the resulting luminosity function $\phi(L)$ has a Schechter form similar to that of the merged CfA1 catalog, and that galaxy luminosity increases monotonically with galaxy mass.

In solving for the "best" luminosity function of the CfA1 catalog, Nolthenius (1993) first assigned distances to each galaxy based on the Burstein-Faber (1987) flow model. Others have used a Virgocentric infall model. Here, for the purpose of comparisons, it is only important that $\phi(L)$ be defined in an equivalent way between the simulations and observations. We therefore assumed simple, unperturbed Hubble flow distances $D = V/H_0$ (except that radial velocities $V < 300$ km s$^{-1}$ are set to $V = 300$ km s$^{-1}$). We next found the Schechter parameters $\alpha$ and $M^*$ for the merged CfA1 catalog using a code based on the inhomogeneity-independent method of deLapparent, Geller & Huchra (1989, DGH). $\phi^*$ was constrained by requiring that the integrated number of galaxies N over the solid angle $A$ of the catalog inside $V = 12000$ km s$^{-1}$ matched that of the CfA1,

$$N = \frac{A\phi^*}{{H_0}^3} \int_{300}^{12000} v^2 \Gamma(\alpha + 1, \lambda(V))\, dv \tag{2}$$

$$\lambda(V) = L_{lim}(V)/L^* = \text{dex}\ \left[.4(M^* - 14.5 + 25 + 5\log(\frac{V}{H_0}))\right]. \tag{3}$$

where $L_{lim}(V)$ is the luminosity of a galaxy of $m = 14.5$ at redshift $V$, and $V/H_0$ is in Mpc. Our $\phi^*$ constraint differs from the maximum liklihood methods (e.g. Efstathiou, *et al.* 1990) or that of DGH, yet reproduces their unmerged CfA1 $\phi^*$'s well. After some experimenting, we found that setting $m = 14.5$ at $V = 300$ km s$^{-1}$ for a halo of $(\delta\rho/\rho)_{cut} = 80$ (CHDM) and 156 (CDM) produced a total density of halos $\sim 0 - 20\%$ above that of the merged CfA1 catalog. We chose $(\delta\rho/\rho)_{cut}$'s of 80 and 156 in order to insure that the same small fractions were culled from both the CHDM and CDM catalogs to bring their number densities in line with the CfA1 catalog. Still, the difference between $(\delta\rho/\rho)_{cut} = 80$ and 156 has only a minor effect on the sky catalogs since the imposition of an apparent magnitude limit means that only a few percent of sky catalog galaxies have $\delta\rho/\rho < 156$. (Contrast this to the situation with the box groups, which are volume limited. The steep mass function then produces a box sample dominated by low mass halos and their resultant groups.) Random culling down to CfA1 density then yielded our final sky catalogs. The 10.384 sr sky catalogs each had 9204 galaxies inside a redshift of $12,000$ km s$^{-1}$. All 2.66 sr CfA-sky catalogs had 2358 galaxies within $12,000$ km s$^{-1}$. Because of the periodic boundary conditions, most bright galaxies appear more than once in the final sky catalog. On average, each galaxy appears $\sim 3-4$ times in the full 10.384 sr sky catalogs, and $\sim 1.8$ times in the 2.66 sr sky catalogs. This would be a concern for any study focusing on large scale structure. However, we focus on median group properties and these should be insensitive to this level of replication.

Our adopted breakup procedure produced a high fraction of fragments in the sky catalogs; 27% (CHDM), 50% (CDM1.5) and 61% (CDM). Using breakup Method 1, a high percentage of sky catalog galaxies were also fragments of overmergers: 27% for CHDM and 70% for CDM. These percentages were only 11% (CHDM), 37% (CDM1.5) and 46% (CDM) for Method 2, which had many more low mass fragments.

The "true" mass of a DM halo is not a well defined concept. The DM around a simulation galaxy merges smoothly with the DM in a group, which merges smoothly with a general background. We believe it's plausible, however, that regardless of how one chooses to define it, the DM mass will have a monotonic relation to our 1-cell mass. If we further assume that DM mass and blue luminosity are monotonically related, then our luminosity assignments should insensitive to the detailed relations, and be reasonably representative. Nevertheless, the relation between DM mass and 1-cell mass is likely to be non-linear, so that one would be suprised if a constant $M/L$ resulted from our luminosity prescription, even if in some sense $M/L$ is a constant for galaxies. Figure 6a shows $M/L$ for all of our sky catalogs. While it diverges upward at the smallest masses (which comprise only a tiny fraction of the sky catalog members), for the great majority of galaxies $M/L \sim 10$ for CHDM and $M/L \sim 20-25$ for CDM. (Other luminosity assignment methods have been tried in earlier dissipationless simulations. For example, Gelb (1992) uses the Tully-Fisher and Faber-Jackson relations. He too finds a luminosity function which is roughly constant over intermediate masses and deviates at the low and high ends.) When the massive halos are broken up, $M/L$ for the fragments ends up lower than for the no breakup case. Figures 6(b) and 6(c) show the distribution of 1-cell masses in the sky catalogs. Note that breaking up the overmergers changes the mass distribution substantially. A strong peak occurs for the fragments, just below the cutoff mass $M_{bu} = 7 \times 10^{11}\ M_{\odot}$ . This is especially true for CDM, which has more and higher mass halos than CHDM.

Table 3 shows the characteristics of the resulting sky catalogs. Note that merging lowers CfA1's luminosity density and brightens $M^*$ compared to the unmerged CfA1, though the effect is very slight with only 44 mergers. This is because new mergers which would ordinarily be formed from galaxies with $m > 14.5$ are missing from the CfA1 sample.

**Table 3. Sky Catalog Parameters**

| Catalog | $V_{pec}^{\dagger}$ | $\alpha$ | $M^*$ |
|---|---|---|---|
| CfA1 (full) | - | -1.27 | -21.08 |
| CfA1 (merged) | - | -1.26 | -21.10 |
| CHDM1 | 852±99 | ~-1.24 | ~-21.23 |
| CHDM2 | 595±247 | ~-1.20 | ~-21.23 |
| CDM1.5 | 700±127 | ~-1.19 | ~-21.19 |
| CDM1 | 991±169 | ~-1.19 | ~-21.19 |

† peculiar velocity of home galaxy and $1\sigma$ range over 6 viewpoints

To test the sensitivity of our comparisons to luminosity assignment, we considered two alternate methods. Both of these alternates were performed on the no breakup cases only. The first was suggested by J. Ostriker (private communication 1993). Cen and Ostriker's (1993; CO) hydrodynamic simulations lead to a relation between CDM total density $\rho_{tot}$ (dark matter + gas/stars) and collapsed baryonic density $\rho_{baryon}$ (presumably stars). If $\hat{\rho} \equiv \rho/\bar{\rho}$ then CO find $\log(\hat{\rho}_{\rm baryon}) = A + B\log(\hat{\rho}_{\rm tot}) + C\log^2(\hat{\rho}_{\rm tot})$. B and C depend mildly on smoothing scale R. In our simulations, we considered $\rho_{tot}$ the density within a 3-cell cube centered on the halo, giving a smoothing scale of R=0.6 Mpc. Extrapolating CO's trends for coefficients B and C down to $R = 0.6$ Mpc then gave $B = 2.3, C = -0.20$. We assume $M_{baryon}/L$ is a constant. The normalization A was set to insure that after magnitude limiting $(\delta\rho/\rho)_{cut} = 80$ galaxie halos to $m = 14.5$ for all simulations, the number of galaxies retained was similar to the merged CfA1 catalog, giving A=6.23 for both CHDM models and A=5.8 for both CDM models. This relation, with our distribution of masses, produces a Schechter-like luminosity function, but with $M^*$ two and a half magnitudes too bright ($\sim -23.5$), and $\alpha$ too steep ($\sim -1.7$). The resulting halo M/L (see Figure 6a), attaches much higher L's to massive galaxies, relative to the Schechter prescription. This leads to a disproportionate number of distant, luminous galaxies at the expense of those closer.

The second method uses the blue Tully-Fisher relation from Fouque, *et al.* (1990). We assume the cell mass M is gravitationally bound so that a circular velocity may be defined

$$V_{circ} = \sqrt{\frac{GM}{r}}. \tag{4}$$

The Fouque, *et al.* Tully-Fisher relation then defines the absolute B magnitude $M_B$ of the galaxy as

$$M_B = -5.5\log(V_{circ}) - \beta \tag{5}$$

where $\beta$ is a calibration. This is equivalent to $M^{1.1}/L =$ const. Using the Fouque, *et al.* value of $\beta = 8.0$ corrected by $-0.37$ for average internal extinction in spirals and by $-0.29$ to put on the Zwicky system results in over an order of magnitude too few galaxies surviving the $m = 14.5$ limit! This isn't too suprising. Our halos are 200 Kpc across, $\sim 10$ times bigger than the visible size of a typical spiral, and the relation between our 1-cell mass $M$ and the observational $M(r \sim 10$ Kpc $)$ is not known. We therefore adopt a $\beta$ which best matches the merged CfA1 galaxy density of 886 sr$^{-1}$. The resulting halo $M/L$ attaches much higher $L$'s to low mass galaxies, relative to the Schechter prescription. This leads to catalogs with a disproportionate number of nearby galaxies and strongly favors fainter, lower $v_{\rm gr}$ groups.

## 7. Tuning the Grouping Algorithm Link Parameters

Conventionally, a group is thought of as a gravitationally bound collection of perhaps 3 to 50 galaxies which has collapsed. For our purposes, we consider a group more generally as any set of galaxies which satisfy a linking criteria on the sky and in redshift. Any valid cosmological model must be able to produce a set of such groups whose typical properties are in good agreement with identically selected real groups *for any link lengths.* It is not necessary that the groups be bound, let alone collapsed or virialized. Nevertheless, it is also of interest to identify that set of groups which is closest to satisfying the conventional definition with the least contamination by interlopers, e.g. for estimating $\Omega$ from the $M/L$ method (see §8). The corresponding link criteria which make these most realistic groups are referred to below as the "fiducial" links.

We use the adaptive grouping algorithm of Nolthenius (1993), which is a modification of the original friends-of-friends algorithm in redshift space given by Huchra and Geller (1982). Galaxies at redshift V km s$^{-1}$ are linked if the separation on the sky is less than $D_L$ and their difference in redshift is less than $V_L$, where

$$R^{sky}_{M.I.S.}(V) = \frac{2}{\pi}\phi^*\Gamma\big(1+\alpha, \lambda(V)\big), \tag{6}$$

$$D_L(V) = D_n R^{sky}_{M.I.S.}(V_0)\left(\frac{\Phi(V_0)}{\Phi(V)}\right)^{1/2}\left(\frac{V_0}{V}\right)^{1/3}, \tag{7}$$

$$V_L(V) = V_5\left(\frac{\Phi(5000\ \mathrm{km\ s^{-1}})}{\Phi(V)}\right)^{1/3}. \tag{8}$$

$\lambda(V)$ is given by equation (3). $V_0$ is an arbitrary redshift distance used for scaling, here taken to be 1000 km s$^{-1}$. $D_n$, an input parameter, is the sky link expressed as a fraction of the sky-projected mean interparticle spacing $R^{sky}_{M.I.S.}$. $\Phi(V)$ is the integrated galaxy luminosity function above the apparent magnitude limit visible at redshift $V$. $D_L$ is designed to insure that a group at the minimum number density contrast just meets the selection criteria at all redshifts (see NW). We scale $V_L$ with the mean interparticle spacing. We chose this scaling, rather than the linear NW scaling (whose resulting groups $\frac{M}{L}(V)$ trend turned out to match slightly better to groups selected using full 3D information at the same overdensity limit), simply because it produced a flatter distribution of group M/L with distance. We have also run cases with the NW version of the algorithm and find negligible differences in group properties and no change in our conclusions. For the purposes of comparing models, either V(L) can be used. We parameterize the size of the link in redshift with $V_5 \equiv V_L(5000\ \mathrm{km\ s^{-1}})$.

Consider a collection of galaxies which are close together in redshift space. Their rms velocities about their mean peculiar velocity will govern how elongated along the line of sight this group appears, providing a diagnostic for determining the optimal redshift link $V_5$. Define the aspect ratio $A_z$ of this "finger of God" seen in redshift space as $A_z = V(\Delta RA + \Delta Dec)/2\Delta V$, where $\Delta$ refers to the maximum extent of the group in right ascension, declination, and line of sight velocity. Define a group's aspect ratio $A_r$ in real space as $A_r = V(\Delta RA + \Delta Dec)/2\Delta(V - V_{pec})$. Note that $A_r$ can be defined only for simulation groups, since the galaxies' true distances are then known. Because groups are often poorly isolated from their neighbors and because of random peculiar velocities, the true depth of groups picked out in redshift space will inevitably be larger than the sky-projected group size; groups will be elongated along the line of sight. This is not the familiar "finger of God" one sees in redshift maps, it is its counterpart seen in real space. This elongation would, on average, disappear if one could assign memberships with perfect knowledge of galaxies' true distances. Figure 7 shows the trend of $A_r$ vs. $V_5$ for our simulations, at the fiducial $D_n = 0.36$. Consider a collection of galaxies all close together in redshift space and with a small rms velocity $v_{\rm gr}$ about their mean peculiar velocity. At $V_5$ below some threshold, one picks out only groups whose members are within $\sim D_L$ of each other in real space, and $A_r$ is constant and near a minimum (in fact, since there are two sky dimensions and only one depth dimension, there is a tendency for $A_r$ to rise slightly at very low $V_5$, when only a fraction of the valid members are being selected). At $V_5$ below the threshold too many valid group members are excluded, and we refer to this as the "clipped regime". Above the threshold $V_5$, $A_r$ begins to rise as outliers begin to significantly contaminate the group. High $V_5$ is then said to produce groups in the "interloper regime". For the CHDM groups, this occurs at $V_5 \sim 300$ km s$^{-1}$. The CDM curves never flatten, showing that groups have significant contamination even at the lowest $V_5$. The CDM1.5 curve slope steepens slightly at $V_5 \sim 500$ km s$^{-1}$ when the rate of contamination rises. The CDM1 curve shows only a very gradual curvature and a slight discontinuity in slope at $V_5 \sim 800$ km s$^{-1}$. These curves suggest fiducial $V_5 \simeq 300$ km s$^{-1}$ for CHDM and $\sim 600-700$ km s$^{-1}$ for CDM. Note that the minimum $A_r$ does not approach 1, due to the inherent loss of depth information in redshift space. It is possible to force rounder groups and thus smaller $A_r$'s by raising $D_n$ and restricting $V_5$, but at the cost of rejecting an unacceptably high number of valid members. For comparison, Figure 7 also shows the curves for groups from a sky catalog made from a 100 Mpc box of Poisson distributed particles with Gaussian random velocities of $\sigma = 200$ km s$^{-1}$ and $\sigma = 350$ km s$^{-1}$. In these catalogs, there is no coherent clustering to confine the depth dimension, which therefore rises steeply with $V_5$ even at small scales. Note by comparison that even the CDM catalogs show significant coherent motion by their more slowly rising curves.

Perhaps the most relevant measure is to ask what $V_5$ is required in order to produce the same median $v_{\rm gr}$ as is seen in groups selected using full 3D information on our fully corrected simulations at the same fiducial $D_n = 0.36$. This is shown in Table 4. This is $V_5 \simeq 400$ km s$^{-1}$ for CHDM, $V_5 \simeq 550$ km s$^{-1}$ for CDM1.5, and 700 km s$^{-1}$ for CDM1. If no breakup is performed, $v_{\rm gr}$ 's are very similar, but require somewhat lower $V_5$'s to match up with 3D groups.

**Table 4. $V_5$ Link Best Matching 3D Groups' $v_{\rm gr}$ 's***

| simulation | $\mathrm{CHDM}_2$ | $\mathrm{CHDM}_1$ | CDM1.5 | CDM1 |
|---|---|---|---|---|
| $V_5$ | 398(371) | 413(330) | 551(456) | 702(575) |
| $\langle v_{\rm gr} \rangle_{med}$ | 126(151) | 130(148) | 187(188) | 241(237) |

$*V_5$ link giving redshift space groups with the same $\langle v_{\rm gr} \rangle_{med}$ as 3D groups, at fiducial $D_n = 0.36$. All in km s$^{-1}$, no breakup case is in parentheses

Finally, we note that the fraction of galaxies grouped $f_{\rm gr}$ will at first rise steeply with $V_5$ as new members are rapidly added from the dense region containing the group. The $f_{\rm gr}$ ($V_5$) curve will then show a distinct drop in slope as it enters the interloper regime, when essentially all valid members have been added and primarily outliers and interlopers from the low density surroundings are incorporated at higher $V_5$. This transition occurs near $V_5 \sim 350$ km s$^{-1}$ for both CHDM simulations, and near $V_5 \sim 600$ km s$^{-1}$ for both CDM simulations (see §9). As these values are representative of the other measures above, we adopt $V_5 = 350$ km s$^{-1}$ as the fiducial redshift link for CHDM, and $V_5 = 600$ km s$^{-1}$ as that for CDM.

## 8. The Dynamical State of the Groups and Implied $\Omega$

To assess the dynamical state of groups, we calculated the number- weighted virial mass estimator for 3D-selected box groups. This is the $M_{vir}$ appropriate if the mass is dominated by a dark matter background distributed like the galaxies, and is given by

$$M_{vir} = \frac{2\sigma r_h}{G}, \tag{9}$$

where, for a group of $n$ galaxies, $\sigma = (n-1)^{-1} \sum_i (v_i - \bar{v})^2$ is the 3D rms velocity about the mean velocity $\bar{v}$ of the group. We then compared this to the true mass $M_{DM}$ in dark particles within the mean harmonic radius $r_h$, where

$$r_h = \frac{n(n-1)}{2 \sum_{i<j}^{n} \frac{1}{r_{ij}}}, \tag{10}$$

and $r_{ij}$ is the separation between galaxies $i$ and $j$.

Figures 8a and 8b show scatter plots of group $\log M_{vir}$ vs. $\log M_{DM}$ for the no breakup and breakup CDM boxes, respectively. Many of the no breakup CDM groups appear to be unbound, with the median $M_{vir}/M_{DM}$ as high as 2.0 for CDM1, while the breakup box appears only slightly biased to high $M_{vir}$. This effect is not due to the higher minimum halo overdensity ($(\delta\rho/\rho)_{cut} = 156$ vs. $(\delta\rho/\rho)_{cut} = 80$ for CHDM), since it is also clearly seen in the $(\delta\rho/\rho)_{cut} = 80$ boxes. It also appears equally strong when only better sampled groups (at least 10 members) are considered. The explanation is that many small groups do not appear in the no breakup cases, because of overmerging. Recall that a halo is defined as a local density maximum, requiring neighboring cells to be relatively underdense. Also, a group must contain at least 3 members to be identified. The overmerged CDM halos often dominate their surroundings to such an extent that too few other local density maxima can be defined, preventing a group from being identified. When these are broken up, new groups result. These groups turn out to have $M_{vir}$ close to $M_{DM}$, on average, so that their inclusion lowers the median $M_{vir}/M_{DM}$ nearer to unity. This issue affects CDM much more than CHDM, since CDM has $\sim 4$ times as many overmergers (see Table 2).

Note that the breakup case in Figure 8b still has these high $M_{vir}/M_{DM}$ groups. Correlating high $M_{vir}/M_{DM}$ in the no breakup CDM box with other group properties, we found that these groups tend to be poorly sampled (all groups with $M_{vir}/M_{DM} > 30$ have 5 or fewer members), have higher than average size, low density contrasts, high peculiar velocities, and tend to lie in denser regions. Using computer visualization and color-coding groups by their $M_{vir}/M_{DM}$ shows these groups tend to be small and lie near the outer parts of dense clumps and filaments. Putting this all together suggests these groups may have experienced tidal shearing sufficient to unbind them. Other recent work supports this idea (e.g. Mamon 1994). Some high $M_{vir}/M_{DM}$'s are no doubt simply due to the inherent high noise in $M_{vir}$. Figure 9c shows $M_{vir}$ vs. $M_{DM}$ for groups from all 6 fiducial linked sky catalogs. Note that the high $V_5$ required to match median 3D $v_{\rm gr}$ 's nevertheless produces a tail of spurious groups with high $v_{\rm gr}$ , little mass within $r_h$, and hence high $M_{vir}/M_{DM}$. The high $V_5$ link essentially produces a floor for $M_{vir}$, which is unreasonably high for the smallest groups. This is a general feature of all group catalogs linked in redshift space (see Geller & Huchra 1983; HG, RGH). Banding is due to using only a 10% sample of the dark particles; the smallest (spurious) groups contain only a few dark particles.

The corresponding plots for the CHDM boxes are in Figure 9. CHDM box groups appear little changed by breakup, partly because they contain only a quarter as many overmergers, and partly because they show lower random velocities on small scales. Note in Figure 9c that the lower $V_5$ level appropriate for CHDM significantly reduces the problem of spurious low $M_{DM}$ but high $M_{vir}/M_{DM}$ groups. Since a low $V_5$ also appears appropriate to the real universe, the problem of contamination and spurious groups may not be as severe as seen e.g. in Geller and Huchra (1983)

Note that massive groups ($M > 10^{14}\ M_\odot$ ) in all boxes appear to define a narrow $M_{vir}/M_{DM}$ which appears clearly virialized, yet which lies systematically higher than the virialization line. There are two biases towards high $M_{vir}/M_{DM}$ which likely account for this, and which are present for all groups. First, the $r_h$ sphere within which particles are counted is centered on the mean position of the halos. Centering instead on the dark particle concentrations lowers $M_{vir}/M_{DM}$. We confirmed this by centering on the mass-weighted mean position of the particle condensations within $1.5r_h$ of the groups and found it raises $M_{DM}$ such that the median $M_{vir}/M_{DM}$ is lowered by 7%. Second, many groups, especially the larger ones, are elongated density enhancements along filaments. Counting within a sphere will include spurious low density areas along the "equator" while missing the high density regions along the "poles". If this bias is $\sim 30\%$ in $M_{DM}$, which seems plausible (see group geometries in the video sequence of Brodbeck, *et al.* 1994), then $M > 10^{14}\ M_\odot$ groups in the breakup boxes appear to obey the virial theorem, on average, in all simulations.

We also determined several measures of the "velocity bias" $b_v$ (Carlberg & Couchman 1989): the velocity of the galaxies compared to that of the underlying cold dark matter particles. The global velocity bias $b_v^{global}$ is defined by a sum over the $i$ galaxies and $j$ DM particles within the box $b_v^{global} = \left(\sum_i v_{i\,gal}^2 / \sum_j v_{j\,DM}^2\right)^{1/2}$.

The results are summarized in Table 5 below and shown in Figure 10. These change significantly when one looks only within the virial radii of the individual groups. The velocity bias within a group's $r_h$ is defined over the $i$ galaxies and $j$ DM particles as $b_v^{grp} = \left(\sum_i (v_i - \bar{v}_{gal})^2 / \sum_j (v_j - \bar{v}_{DM})^2\right)^{1/2}$, where $\bar{v}_{gal}$ is the mean unweighted galaxy velocity and $\bar{v}_{DM}$ is the mean DM particle velocity within the group. CHDM galaxies show a stronger bias within groups. CDM1 galaxies, however, show little or no bias, in or out of groups, in agreement with $b_v^{global} = 0.94$ from Katz, Hernquist, & Weinberg's (1992) hydrodynamic simulation. CDM1.5 shows a moderate bias both globally and inside groups.

**Table 5. Velocity Bias**

| simulation | $b_v^{global}$ | $< b_v^{grp} >_{avg}$ | $< b_v^{grp} >_{med}$ |
|---|---|---|---|
| $CHDM_1$ | 0.93(0.90) | 0.76(0.78) | 0.78(0.76) |
| $CHDM_2$ | 0.92(0.89) | 0.72(0.69) | 0.72(0.71) |
| CDM1.5 | 0.83(0.79) | 0.76(0.74) | 0.75(0.79) |
| CDM1 | 0.95(0.90) | 0.97(0.92) | 0.98(1.06) |

No breakup case is in parentheses

The standard method for estimating Ω from bound virialized groups is to assume all galaxies have the same mass-to-light ratio M/L, given by the median M/L for groups, then integrating over the luminosity function to get the mass density (Kirschner, Oemler & Schechter 1979). For a Schechter luminosity function this means

$$\Omega = \frac{8\pi G}{3H_0^2}\langle\frac{M}{L}\rangle\phi^* L^* \Gamma(\alpha+2). \tag{11}$$

Table 6 shows the resulting inferred Ω using the fiducial grouping algorithm parameters and Schechter L(M) for the sky catalogs. Means and standard deviations are over the six viewpoints. Using the Cen-Ostriker L(M) gives very similar Ω's. No N(z) corrections (see §9) were applied to get $< M/L >_{med}$ since group $M/L$ is virtually independent of redshift.

**Table 6. Ω from Group M/L Method**

| catalog | $V_5$ | $< M/L >_{med}$ | Ω |
|---|---|---|---|
| CfA1 (full)* | 350 | 62 | 0.06 |
| CfA1 (merged) | 350 | 79 | 0.08 |
| $CHDM_1$ | 350 | $118 \pm 19 (127 \pm 9)$ | $0.12 \pm .014 (0.12 \pm .003)$ |
| $CHDM_2$ | 350 | $139 \pm 12 (120 \pm 7)$ | $0.12 \pm .010 (0.11 \pm .003)$ |
| CDM1.5 | 350 | $159 \pm 9 (164 \pm 7)$ | $0.15 \pm .007 (0.16 \pm .009)$ |
| CDM1 | 350 | $177 \pm 9 (186 \pm 20)$ | $0.18 \pm .007 (0.19 \pm .012)$ |
| | | | |
| CfA1 (full)* | 600 | 119 | 0.12 |
| CfA1 (merged) | 600 | 152 | 0.15 |
| $CHDM_1$ | 600 | $198 \pm 14 (193 \pm 10)$ | $0.21 \pm .008 (0.18 \pm .008)$ |
| $CHDM_2$ | 600 | $200 \pm 14 (178 \pm 10)$ | $0.18 \pm .010 (0.16 \pm .007)$ |
| CDM1.5 | 600 | $350 \pm 23 (308 \pm 16)$ | $0.34 \pm .030 (0.31 \pm .016)$ |
| CDM1 | 600 | $365 \pm 24 (372 \pm 26)$ | $0.37 \pm .012 (0.38 \pm .022)$ |

* full, unmerged CfA1 catalog.

* No-breakup results in parentheses

Note that all of our $\Omega = 1$ simulations yield low observed $\Omega$'s. Thus, these simulations show a feature which has been seen in real data for over a decade and which has been used by some to argue for a low $\Omega$ universe: How can $\Omega$ be 1 if bound groups and clusters consistently account for only $\simeq 10 - 20\%$ of the mass? Three factors can explain this discrepancy. (1) Only the mass within the mean harmonic radius $r_h$ of the groups is measured by the virial estimator. We saw in Figure 3 that density falls approximately like an isothermal sphere outside group cores, and that this continues to at least $2r_h$ (about midway to the next nearest group, on average), where the cumulative mass $M(2r_h) = 3M(r_h)$ (see Figure 4b). This additional factor of 3 in mass is almost certainly bound to the groups, since $\rho(2r_h) \simeq 4\rho_c$(CHDM) to $10\rho_c$(CDM), and infall is still occurring. The fraction of the total mass in the box which lies within the $r_h$ sphere of fiducial linked groups is only $\sim 15 - 25\%$, as shown in Table 7. Most mass lies outside of groups.

**Table 7. Mass, Volume Fraction Within Box Groups**

| simulation | $N^*_{grps}$ | $f^{**}_{mass}$ | $f^{\dagger}_{volume}$ |
|---|---|---|---|
| $\mathrm{CHDM}_1$ | 575 | 0.14 | 0.0086 |
| $\mathrm{CHDM}_2$ | 640 | 0.15 | 0.0072 |
| CDM1.5 | 736 | 0.23 | 0.0038 |
| CDM1 | 617 | 0.27 | 0.0026 |

* Number of groups in box

** fraction of box mass which is within $r_h$ of groups

† fraction of box volume which is within $r_h$ of groups

(2) The relevant virial velocities appropriate for measuring the local mass are actually those of the individual particles, not the galaxies. As already seen in Table 5 and Figure 10, the median velocity bias within CHDM groups is $\sim 0.75$. Since $M_{true}/M_{vir} \propto 1/b_v^2$, this contributes another factor of 1.8 to the $\Omega$ estimate. (3) We've also assumed spherical symmetry in Figures 3 and 4. The true shape of groups is more elongated, as structure is still quite stringy at this point in the evolution. The inappropriately counted matter along the "equator" of the groups is of lower density and fails to compensate for the uncounted high density regions missing above the "poles" by perhaps a factor of 1.3. The product of these factors ($\sim 7$) approaches an order of magnitude. While factor (1) above is closely related to the fact that the galaxies are a biased (b=1.5) tracer of the DM, it seems likely that the less clustered hot DM, especially that filling the voids (see Brodbeck *et al.* 1994) will additionally bias $M/L$ to the low side. Together, these factors show that, within the assumptions of this method, an observed $\Omega \simeq 0.1$ can indeed be consistent with a true $\Omega = 1$ universe. In fact, virial estimates appear to be underestimates of the true mass in real clusters as well. At present, gravitational lensing appears to be the most reliable method of measuring total masses, and several studies show that typically the resulting total masses of clusters out to just beyond the optical radii is roughly a factor of 2 or 3 higher than virial estimates (e.g. Kaiser, *et al.* 1994). There are other, more subtle problems with the M/L method; for example, the stellar populations of groups and especially clusters is well known to be older and have lower M/L than is typical for the field galaxy populations.

Our $\Omega$ at $V_5 = 350$ for the CHDM simulations is 50% higher than that for the merged CfA1 at the same $V_5$: $\Omega_{CHDM} = 0.12$ vs. $\Omega_{CfA1} = 0.08$. The difference is due almost entirely to the difference in the median of the mean harmonic radii of groups, as discussed in §10. Our $\Omega$ for the full, unmerged CfA1 is smaller than that of HG's $\Omega \simeq 0.1$ for the nearby, all sky m=13.2 CfA sample, and smaller than RGH's $\Omega \simeq 0.13$. As we've argued elsewhere (NW, N93), we prefer a significantly smaller redshift link than these authors, which directly lowers the virial masses. The RGH work is for the CfA Slices, whose group's show a suprising 60% higher $v_{\rm gr}$ than for their equivalently selected CfA1 groups, due to differing sample depth and, to some extent, cosmic variance (see RGH for discussion). Also, their link parameters differ from ours. They used a smaller sky link (giving smaller $r_h$) but larger redshift link, leading to higher $v_{\rm gr}$ 's and net higher $M/L$'s.

When using full 3D information to make sky catalog groups ($D_n$=0.36 as before), our median $\Omega$'s are higher: $\Omega$ =0.17(both CHDM), 0.36(CDM1.5), and 0.48(CDM1). Note that the $V_5's$ necessary to match 3D $v_{\rm gr}$ 's are, except for CDM1.5, higher than our fiducial $V_5$, as shown in Table 4. For example, raising CHDM's $V_5$ from the fiducial 350 to 398 km s$^{-1}$ raises $< v_{\rm gr} >_{med}$ by $\sim$ 15% and hence $< M/L >_{med}$ and $\Omega$ by $\sim$ 32%, lessening the difference between the observationally inferred and true $\Omega$. The same holds for CDM1. Our CDM1.5 estimate of $\Omega \simeq 0.33$ agrees well with that of Katz, Hernquist & Weinberg's (1992) TREESPH hydrodynamic simulation result of $\Omega \simeq 0.31$ on a much smaller sample.

## 9. Sky Catalog Results and Comparisons with CfA1

When comparing our sky catalog groups with those of CfA1, there is one more calibration to consider. To maximize the amount of precious observational data used, we retain the full magnitude limited CfA1 catalog rather than a volume limited subset. In a magnitude-limited catalog, both the sky and redshift "friends-of-friends" links scale up with distance, and group properties will change significantly with distance. In particular, more distant groups will be larger, brighter, and have higher $v_{\rm gr}$ 's. Median values will therefore be biased if groups are distributed differently with distance between the two datasets. Chance differences in large scale structure will mean, in general, that groups are in fact distributed differently with redshift. Our home galaxy selection criteria do not fully insure that simulation groups are distributed in redshift like CfA1 groups. Equally important is our small box size. Periodic boundary conditions give repeating structures every 100 Mpc = 5000 km s$^{-1}$, whereas the CfA1 data are actually rather sparse beyond the Virgo Cluster. Figure 11 shows the galaxy density vs. redshift for the merged CfA1 and CHDM$_2$ sky catalogs. Relative to the CfA1 dataset, CHDM$_2$ remains underpopulated out to $\sim$ 3000 km s$^{-1}$ and overpopulated beyond. The other simulations follow this same pattern. Not correcting for this "N(z) bias" will lead to overestimating the average or median sizes and $v_{\rm gr}$ 's of simulation groups, as well as affect other properties. We've handled this by averaging results for four random subsamples of the simulation groups such that, when their redshifts are binned to 1000 km s$^{-1}$ bins, the number of groups vs. redshift N(z) matches that of the CfA1 groups selected at the same links. The median group properties we present below which are medians from these subsamples are labelled "N(z) corrected" for clarity.

In Figure 12 we show $v_{\rm gr}$ vs. $V_5$ at our fiducial $D_n$ = 0.36. It is analagous to Figure 2 in NKP94, which was done at $D_n$ = 0.47 to show good agreement in CDM1.5 velocities between our work and that of Moore, Frenk and White (1993; MFW). They used a quite different particle-particle-particle-mesh code with better spatial resolution but poorer mass resolution than ours here. This figure emphasizes again one of our main conclusions; CDM groups have much higher internal rms velocities than observed, while CHDM is in good agreement. It may be suprising to see that the curve for CDM1.5 differs so little from that of CDM1, since numerous studies show biased CDM has lower velocities. Our global rms peculiar velocities (i.e. for all galaxies within the box) are indeed much higher for CDM1 than those for CDM1.5; 944 km s$^{-1}$ vs. 650 km s$^{-1}$, or 45% higher. Within groups, however, two different effects combine to reduce this difference. First, on small scales ($\simeq$ 1 Mpc), CDM1 rms peculiar velocities are only 30% higher than for CDM1.5. This is true both for the full set of box groups, and for the 3D selected groups from the sky catalogs. Second, in redshift space, the size of the $V_5$ link strongly affects $v_{\rm gr}$ . Using the same $V_5$ = 600 km s$^{-1}$ on both simulations will then further reduce their differences to only 7% ($v_{\rm gr}$ = 216 km s$^{-1}$ vs. 202 km s$^{-1}$) This same effect is seen in MFW, and our result agrees well with the 8% difference seen between their b=1.6 and b=2.0 CDM median $v_{\rm gr}$ 's.

It is important to note that the error bars on these and later curves are $1\sigma$ deviations from the 6 different sky catalogs. While our intent was to measure cosmic variance, in fact these are a significant underestimate. Because of the small size of our box, many of the same groups are seen in most or all viewpoints, (albeit sampled differently due to the magnitude limit). The difference between the CHDM$_1$ and CHDM$_2$ curves is a fairer estimate of cosmic variance. Our estimated probability of a 100 Mpc $^3$ box having the power spectrum of CHDM$_1$ of $\sim$ 10% (NKP) corresponds to about $1.7\sigma$. If so, then the difference between the CHDM$_1$ and CHDM$_2$ curves may be a rough estimate of $1.7\sigma$ error bars due to cosmic variance. This is the method assumed in quantifying comparisons with CfA1 below.

Another important point is that since CHDM$_1$, CDM1, and CDM1.5 all have higher power in the longest waves within the box, their "Virgo"'s are larger and richer, and the fraction of galaxies in groups is artificially higher than would be typical. If one believed the CfA1 data were a "fair sample" (but see below), then comparing to CHDM$_2$ would be more appropriate. To estimate this effect on the CDM curves one can do a rough calibration by looking at the CHDM$_2$ curves shown here and below and shift the CHDM$_1$ curve towards it, carrying along rigidly the two CDM curves in the same direction. In fact, however, the luminosity density of the CfA1 sample appears to be $\sim$ 25% higher than for the much larger (fair sample?) APM data (Tully, private communication). If the presense of rich structures like Virgo, Coma, and the Great Wall in this sample similarly indicates slightly unusual higher power on larger scales, then the appropriate curve to compare to CfA1 may be intermediate between CHDM$_1$ and CHDM$_2$.

Figure 13 shows $f_{\rm gr}$ , the fraction of galaxies grouped, vs. $V_5$. Fraction grouped is a powerful statistic, since by separating our grouping algorithm into velocity and sky projected components, $f_{\rm gr}$ is sensitive not only to small scale pairwise velocity differences, but also the degree of spatial concentration of galaxies. CDM differs from CHDM significantly on both measures. As remarked earlier, the improved merging scheme applied to the CfA1 data here had the effect of performing fewer mergers. This left more CfA1 members in dense regions and thus raised the fraction of galaxies grouped; e.g. from 42% to 49% at the fiducial links. It had very little effect on the $v_{\rm gr}$ . Another change from NKP94 was to make CDM sky catalogs using only halos above $(\delta\rho/\rho)_{cut} = 156$ rather than 80. This had the effect of slightly raising CDM's $v_{\rm gr}$ 's by $\sim$ 3%. The net result is closer agreement between CHDM$_2$ and CfA1 $f_{\rm gr}$ curves. CHDM$_1$'s higher power on large scales leads to a significantly higher correlation and a too high $f_{\rm gr}$ . CDM's high small scale pairwise velocities and puffier, less concentrated galaxy filaments inhibit grouping, especially near the fiducial links.

By combining Figures 12 and 13 to eliminate the grouping link $V_5$, we can both enhance the differences between CDM and CHDM and produce a statistic which is very robust. This, our favored statistic, $v_{\rm gr}$ vs. $f_{\rm gr}$ , is shown in Figure 14. CHDM$_2$ is in close agreement with the CfA1 data. CHDM$_1$ groups too high a fraction of galaxies, while CDM both groups too few galaxies and produces $v_{\rm gr}$ 's too high, at the $\sim 4\sigma$ level (due to cosmic variance) for CDM1 and $\sim 5\sigma$ level for CDM1.5. The suprisingly high discrimination shown by this statistic shows the power of grouping in redshift space. The "pressure" provided by higher small scale pairwise velocities will tend to expand spatial structures on $\sim$Mpc (i.e. galaxy group) length scales. This not only lowers the fraction of galaxies grouped, but also raises their rms velocities, and does so at all link criteria. $f_{\rm gr}$ is sensitive to the presense of large clusters in relatively small samples like the CfA1 data. It is therefore important to see how $v_{\rm gr}$ vs. $f_{\rm gr}$ behaves on simulation data with the same sky coverage as CfA1, shown in Figure 14(b). The dominance of the local "Virgo"'s tends to raise $f_{\rm gr}$ for the CHDM$_1$, CDM1, and CDM1.5 simulations (which, recall, have not only higher power, but identical locations for large scale structure). CHDM$_2$, which has smaller "Virgo"'s, actually shows a small decrease in $f_{\rm gr}$ . Median $v_{\rm gr}$ , as expected, shows little change. The net conclusion remains the same, albeit noisier and with higher cosmic variance. $CHDM_2$ again shows excellent agreement with CfA1 throughout the $V_5$ range. Comparing data point by point, we see that $CHDM_1$ has $v_{gr}$ 's slightly too high, while $CHDM_2$'s agree closely with observations. $CHDM_1$ now seems clearly to group too many galaxies. For a curve intermediate between $CHDM_1$ and $CHDM_2$ (see §10) Figure 14b can then be interpreted as indicating that group analysis favors a slightly lower $\Omega_\nu$ than the 0.30 used here.

The $f_{gr}$ vs. $v_{gr}$ statistic is also quite robust. Figure 15 shows $v_{gr}$ vs. $f_{gr}$ for the no breakup simulation sky catalogs, without N(z) correction, for our three L(M) assignment methods. Using the (b) Cen-Ostriker L(M) prescription leaves the curves almost unchanged, even though the luminosity function's $M^*$ is now $\sim 2.5$ magnitudes too bright and the faint end slope is $\alpha \sim -1.8$. The (c) Tully-Fisher (TF) prescription's L(M) differs even more drastically from observations, and is quite un-Schechter-like. Still, the resulting $v_{gr}$ vs. $f_{gr}$ curves are qualitatively similar to the Schechter and Cen-Ostriker results, and again CDM curves are too high while CHDM's curves are now too low. Note that all simulation TF $v_{gr}$ 's are lower. This is because the TF L(M) strongly picks out nearby, low-mass galaxies and groups and therefore gives lower median $v_{gr}$ 's. Comparing Figures 14a and 15a shows that breaking up the halos and correcting for differing redshift distribution lowers $v_{gr}$ , but still leaves CDM much too high and $CHDM_2$ in good agreement with observations. Notice also that the corrected CDM1.5's curve actually lies above the CDM1 curve. The reason is that CDM1.5 groups a significantly lower fraction than CDM1. Thus it is more accurate to say that the CDM1.5 curve is *left* of the CDM1 curve.

Figure 16 compares the breakup methods on the $v_{gr}$ vs. $f_{gr}$ plane. Relatively little change is seen for any method when applied to CHDM, as shown in Figure 16a for $CHDM_2$. For CDM1, shown in Figure 16b, all methods are close except for breakup Method 1. By forcing essentially equal, maximum possible masses for all DM halo fragments, Method 1 raises $f_{gr}$ dramatically while lowering $v_{gr}$ . Figures 16a and 16b Method 1 curves lie virtually on top of one another. In fact, breakup Method 1 has the unfortunate property that all simulation curves are almost degenerate on this plane, and significantly below the CfA1 curve. While there is no reason to think that this breakup scheme is reasonable, it does show that it is possible to construct sky catalogs whose $v_{gr}$ vs. $f_{gr}$ properties are not discriminated.

Figures 17-19 show how $v_{gr}$ vs. $f_{gr}$ change when $V_5$ is held fixed and $D_n$ is varied instead. Here, groups grow primarily on the sky and only secondarily in redshift depth. In NKP we described Figure 17, which does not have the breakup or N(z) corrections included. Our conclusion there was that the dense cores picked out at small $D_n$ were significantly "cooler" in CfA1 than for any simulation. Figure 18 shows that this is only true when overmergers are not broken up and no N(z) correction is made for differing redshift distributions. When corrected in this way, CHDM actually reproduces the CfA1 results very well, while CDM $v_{gr}$ 's still remain too high. The largest effect is the N(z) correction. As described earlier, the simulation groups tend to lie at higher distance, where the magnitude limit then identifies larger, higher $v_{gr}$ groups. Taking the median of random sub-samples whose distribution in distance matches that of the CfA1 lowers median $v_{gr}$ 's. Comparing Figures 17 and 18 shows that $f_{gr}$ in fact changed very little at these fiducial $V_5$'s, while most of the change is in $v_{gr}$ . Figure 19 is similar to Figure 18, but for the CfA-sky catalogs. Again, at these low and moderate $V_5$'s, $f_{gr}$ changes very little between these samples.

Figures 20 and 21 shows the number of groups per steradian vs. redshift link, and vs. sky link, and demonstrate the percolation properties of the catalogs. As $V_5$ is raised, new groups will be identified, and some existing groups which are close in redshift will be merged. Beyond $V_5 = 350$ km s$^{-1}$, the $CHDM_1$ curve actually shows a significant decline as merging outpaces the production of new groups. An intermediate curve between the two CHDM curves would appear to agree much better with CfA1 groups, but appears likely to peak at both $D_n$ and $V_5$ too low. Comparing such a CHDM curve to the CDM curves, which roughly peak together with CfA1, indicates that a lower $\Omega_\nu$ would improve CHDM's agreement. Pure CDM, however, appears to significantly underproduce groups at small $V_5$. Thus, on this measure, CHDM appears to agree well if CfA1 has a moderately high amount of large scale power, and if we lower $\Omega_\nu$ (but not to zero).

## 10. Possible Problems

One quantity which shows significant disagreement with observations for all simulations is group size. Figure 22 shows the median $r_h$ vs. $D_n$. All simulations are $50\% - 80\%$ higher than CfA1, at least near the fiducial link of $D_n = 0.36$. CDM actually fits slightly better. This is because CDM groups galaxies less efficiently and groups tend to be poorer and smaller than CfA1 groups. We've been careful to filter the observations to the same spatial resolution as the simulations, so the difference may be real. Or, it may instead reflect fundamental limitations in our simulations. One possibility is that the overdense DM cells we call galaxies still retain some of the distribution properties of their parent DM, and are insufficiently "galaxy- like". Serna, *et al.* (1994) show that cluster galaxies residing in a DM background will show a stronger concentration than the DM, by a factor of about $r_h(DM)/r_h(gal) \simeq 2$, higher for older clusters. If CfA1 galaxies are like Serna's galaxies, while our DM cells behave somewhere between galaxies and DM particles, one might expect $r_h(gal)/r_h(DM)$ values similar to what we see.

Interestingly, our (disfavored) breakup Method 1 gave median $r_h$ values which were closer to that for CfA1. This is because such a large number of fragments met the magnitude limit, since we artificially imposed nearly equal masses on all fragments. This produces many more very small groups. Even so, the trend of median $r_h$ with $D_n$ was still steeply negative, similar to Figure 22. Recall that with Method 1 all simulations produced virtually the same $v_{\rm gr}$ vs. $f_{\rm gr}$ curves and in poor agreement with CfA1.

## 11. Discussion and Conclusions

We've shown that COBE-normalized CHDM at $\Omega_\nu = 0.3$ produces group properties similar to those of the CfA1, while CDM at b=1.0 and b=1.5 groups too few galaxies and gives $v_{\rm gr}$ too high, at the several $\sigma$ level. We now attempt to refine our estimate for an $\Omega_\nu$ which is in optimal agreement with our group analysis.

The $\mathrm{CHDM}_1$ power spectrum in real space $P_r(k)$ is a factor of 2 higher than that for $\mathrm{CHDM}_2$ on scales comparable to the size of the box (KNP). The redshift space power $P_z(k)$ is amplified by a factor

$$P_z(k) = P_r(k)\left[1 + \frac{2}{3}\frac{\Omega^{0.6}}{b} + \frac{1}{5}\left(\frac{\Omega^{0.6}}{b}\right)^2\right] \qquad (12)$$

(Kaiser 1987). For our $\Omega = 1$ b=1.5 CHDM simulations then, $P_z(k) = 1.5P_r(k)$, so that in redshift space $\mathrm{CHDM}_1$ has $P_z(k)$ 3 times higher than that of $\mathrm{CHDM}_2$. The CfA1 $P_z(k)$ is approximately a factor of 1.6 higher than that of the CfA2 at these scales (Vogeley, *et al.* 1992), while it may be more comparable to CfA2 on smaller scales. The CfA2 $P_z(k)$ is in turn a factor of 2 higher than that of the much larger APM Survey on similar scales (Baugh & Efstathiou 1993). While these scales are larger than typical group/clustering lengths, our $\mathrm{CHDM}_1$ vs. $\mathrm{CHDM}_2$ simulations show that higher power on these scales indeed "crosstalks" into higher fractions of galaxies grouped (see Figure 13), perhaps aided by percolation along extended filaments. Such coupling of large to small scales has already been noted for pairwise velocities (Gelb, *et al.* 1993). Also, on galaxy scales the CfA1 galaxy luminosity density is a factor of 1.25 higher than that of the APM. If the APM can be taken as approaching a fair sample of the universe, the CfA1 then appears to have a redshift space power spectrum at large scales which is at least twice as high as is typical. This then suggests that the proper curve to compare to the CfA1 data is intermediate between those for $\mathrm{CHDM}_1$ and $\mathrm{CHDM}_2$, but closer to $\mathrm{CHDM}_1$. Figures 13 and 14 then indicate an optimum $\Omega_\nu$ somewhat lower than 0.3. While the detailed relation between $\Omega_\nu$ and $v_{\rm gr}$ vs. $f_{\rm gr}$ is as yet unexplored, if it is approximately linear then Figures 13 and 14 suggest an optimum $\Omega_\nu \simeq 0.2$. With the intermediate CHDM curve described above, Figures 20 and 21 would be in excellent agreement with the CfA1 data for any $\Omega_\nu$ significantly less than 0.3. Figures 18 and 19 favor $\Omega_\nu \sim 0.3$, but are, within the errors, compatible with a slightly lower value. The total mass $m_\nu$ of all massive neutrino species is related to $\Omega_\nu$ by $m_\nu = 23.51\Omega_\nu h_{50}^2$ eV, where $h_{50}$ is the Hubble parameter in units of 50 km $\mathrm{s}^{-1}\mathrm{Mpc}^{-1}$, using the current CMB temperature of 2.726 K, and relevant parameters from Kolb & Turner (1990). The present group analysis' favored $\Omega_\nu \sim 0.20$ then corresponds to $m_\nu \simeq 4.6$ eV, for $H_0 = 50$ km $\mathrm{s}^{-1}\mathrm{Mpc}^{-1}$.

The void probability function is perhaps another indication that CHDM works better for lower $\Omega_\nu$ (Ghigna *et al.* 1994). Also, CHDM forms structure so late that observations of damped Lyman alpha systems in quasars are now only marginally consistent with the CHDM parameters studied here (Klypin, *et al.* 1994a and references therein). However, the abundance of massive objects at high redshift is quite sensitive to $\Omega_\nu$, since these objects are far out on the exponential tail of the mass distrubution. Reducing $\Omega_\nu$ to 0.20 or even 0.25 brings good agreement with current observations (Klypin, *et al.* 1994a).

The second major conclusion of the present paper is that, within the models studied here, it is quite natural to find $\Omega \sim 0.1$ from group mass to light ratios, even though the true $\Omega$ is 1. Three factors combine in the same direction to severely bias the M/L method on the low side. First, galaxies within a group occupy only the central core of much larger DM concentrations. The total mass bound to the group at $2r_h$ is a factor of 3 higher than the DM within $r_h$. Second, the appropriate virial mass to calculate is that due to the individual cold DM particles, not the galaxies, whose velocity bias contributes another factor of $\sim 2$. Finally, the virial theorem implicitly assumes spherical symmetry, and our groups are actually fairly elongated. Counting mass within these elongated boundaries will add perhaps another 30% to the true mass, giving a net correction to $M/L$ of nearly an order of magnitude. Hot DM which is unclustered will further bias $M/L$ to the low side.

## 12. Acknowledgements

It is a pleasure to acknowledge stimulating and sometimes spirited discussions with Neal Katz and David Weinberg, and to thank Gus Evrard for suggestions on how to break up overmergers. We also thank Jerry Ostriker for suggesting the Cen-Ostriker work as a way to relate mass to luminosity. RN gratefully acknowledges a NASA grant administered by the American Astronomical Society, and a continuing grant of computer resources by UCO/Lick Observatory. AK and JRP acknowledge support from NSF grants. The simulations were run on the Convex C-3880 at the NCSA, Champaign-Urbana, IL.

Figure Captions

Figure 1. Comparison of the two methods of assigning velocities to the fragments of (a) CHDM and (b) CDM1 overmerger breakups. $V^{om}_{neigh}$, the rms velocity of all neighboring galaxies within 1 Mpc , or within whatever larger radius encloses 4 neighbors, has a much wider range than the characteristic circular velocity $V^{om}_c = (GM_{eff}/r_{eff})^{1/2}$, yet the medians are very similar. CDM1 has a wider range for both velocities than does CHDM.

Figure 2. Comparison of the halo 1-cell mass distribution for the different massive halo overmerger breakup methods. Our favored method is labeled "breakup", while Methods 1 and 2 are labelled "breakup$_1$" and "breakup$_2$". d$LogN$/d$logM$ $= -1.3$ below the breakup mass $M_{bu}$.

Figure 3. The DM density distribution for (a) the CHDM simulation box groups made from halos with $\delta\rho/\rho > 80$, and (b) the CDM1 and CDM1.5 box groups made from halos with $\delta\rho/\rho > 156$, compared to that for $r^{-2}$ isothermal spheres. r=0 is the uweighted average of the position vectors of all member halos. Only the curve for total mass is shown for CHDM$_1$, while CHDM$_2$ shows the hot and cold distributions as well. Even at $r = 2r_h$, mass is substantially above the critical density $\rho_c$. All simulations show an exponential $\rho(r)$, but CDM shows a steeper fall-off than CHDM. Data inside $\sim 195kpc$ is unresolved and not plotted.

Figure 4. The cumulative mass distribution around (a) the fiducial box groups and (b) around the fiducial sky groups. There is substantial mass outside but bound to the box groups; $M(2_rh)/M(r_h)$ is 2.7 for CHDM and nearly 2.0 for CDM. Around the sky groups, $M(2_rh)/M(r_h)$ is as high as 3.0 for all simulations. Blurring in redshift space reduces the differences between CHDM and CDM curves in the sky groups. Data inside $\sim 195kpc$ is unresolved and not plotted.

Figure 5. The ratio of cold to hot dark matter for the CHDM simulation box groups. This is an unweighted average over all groups. It is more representative of small groups, which make up most of this volume-limited sample. Group cores are relatively dense and cold, with the hot particles predominately forming a more diffuse background.

Figure 6. (a) The assigned 1-cell mass-to-light ratio required in order to reproduce the CfA1 Schechter luminosity function parameters, for each simulation. There is only a single curve for the Cen-Ostriker and Tully-Fisher prescriptions, since no attempt is made to force the resulting luminosity function to agree with observations. The steep rise in M/L at low mass means low mass galaxies do not appear in the sky catalogs. (b) The distribution of no breakup sky catalog galaxies vs. mass. (c) Breaking up overmergers clips the high end and makes a large number of fragments below $M_{bu}$. Tully-Fisher (not shown) and Cen-Ostriker L(M)'s put proportionately much higher luminosity on faint, low mass, nearby galaxies.

Figure 7. The median axial ratio (depth divided by average of RA and Dec dimensions) in real space of redshift-selected groups, vs. $V_5$. The CHDM curves flatten at low $V_5$, indicating minimal contamination by interlopers here. CDM curves are steeper and more contaminated. For comparison, Poisson distributed galaxies of rms peculiar velocities $\sigma = 200$ and 350 km s$^{-1}$ are much steeper still, showing even CDM galaxies have significant coherent motion.

Figure 8. The number-weighted virial mass estimator $M_{vir}$ from the halos, vs. the dark matter $M_{DM}$ contained within the mean harmonic radius $r_h$ defined by the halos, for 3D selected box groups in CDM1. The no-breakup case (a) has a significant tail of high $M_{vir}/M_{DM}$ groups, while the breakup case (b) includes many more $M_{vir}/M_{DM} \sim 1$ groups. Higher pairwise velocities in CDM cause the extended tail to high $M_{vir}/M_{DM}$ in both cases. The combined sky groups for all six viewpoints (c) show a strong bias towards high $M_{vir}/M_{DM}$ at low mass, as high pairwise velocities mean small sky groups are not bound.

Figure 9. The same as Figure 9, but for $\mathrm{CHDM}_2$. There is little difference between the (a) no-breakup and (b) breakup cases, and the low pairwise velocities make for less bias and less noise in the (c) sky groups' $M_{vir}$ than seen in CDM.

Figure 10. The velocity bias parameter $b_v$ for all box groups at the fiducial link. The horizontal lines define the median values. $\mathrm{CHDM}_2$ has a significant bias of $b_v = 0.7$. CDM1 shows almost no net velocity bias. $\mathrm{CHDM}_2$'s low pairwise velocities lead to a tighter distribution. The banding in CDM1.5 (d) is due to the small subsample of dark particles used in the calculations.

Figure 11. The galaxy density vs. redshift distribution in the magnitude limited sky catalogs for $\mathrm{CHDM}_2$ vs. that for CfA1. Since the simulations' densities drops less steeply with redshift due to the limited box size and cosmic variance, requiring random sub-samples of simulation groups to match CfA1's group distribution N(z) insures that median properties have no distance-dependent bias. This is the N(z) correction.

Figure 12. Median $v_{\rm gr}$ vs. $V_5$, for sky catalogs from the breakup boxes and the CfA1. CHDM curves match the observations well, while CDM's are too high. Below $\sim 400$ km s$^{-1}$, curves are degenerate since the CDM groups are in the "clipped" regime. On this and later curves, error bars are $1\sigma$ over the 6 viewpoints.

Figure 13. $f_{\rm gr}$ vs. $V_5$ for all simulations. CDM groups too few galaxies. $\mathrm{CHDM}_1$'s higher power on large scales leads to grouping too high a fraction, while $\mathrm{CHDM}_2$ agrees well with observations.

Figure 14. Our favored statistic $v_{\rm gr}$ vs. $f_{\rm gr}$ under varying $V_5$ for our fully corrected case using (a) the 10.384 sr sky catalogs and (b) the 2.66 sr CfA-sky catalogs. $\mathrm{CHDM}_2$ is in close agreement with the CfA1 data. CDM curves in (a) are too high by an estimated $\sim 6\sigma$ due to cosmic variance, assuming the difference between $\mathrm{CHDM}_1$ and $\mathrm{CHDM}_2$ is $\sim 1.7\sigma$.

Figure 15. Comparison of $v_{\rm gr}$ vs. $f_{\rm gr}$ for the fiducial (a) Schechter, (b) Cen-Ostriker, and (c) Tully Fisher luminosity assignment methods. Comparisons were done before breaking up overmergers or applying an N(z) correction, and the purpose here is mainly to demonstrate robustness. In all cases, CDM is substantially too high. The Tully-Fisher case is probably the least reliable method, as it produces a luminosity function which is quite un-Schechter-like. All methods give similar conclusions, though TF predicts a lower $\Omega_\nu$ than the others.

Figure 16. Comparison of $v_{\rm gr}$ vs. $f_{\rm gr}$ for different overmerger breakup methods for (a) $\mathrm{CHDM}_2$ and (b) CDM1. Other simulations are similar. All methods except Method 1 all lead to CDM curves several $\sigma$ too high. Method 1 collapses all simulations onto the same curve (e.g. (a) and (b) dotted curves virtually overlap) by forcing all fragments to have the maximum possible nearly equal brightness.

Figure 17. Median $v_{\rm gr}$ vs. $f_{\rm gr}$ when varying $D_n$ and holding $V_5$ constant at the two different fiducial values (a) $V_5 = 600$ and (b) $V_5 = 350$ km s$^{-1}$. These comparisons were done with no breakup of overmergers and no N(z) correction. $v_{\rm gr}$ curves are fairly flat, since groups grow mostly on the sky direction when varying $D_n$. At $V_5 = 350$ the CDM groups are in the "clipped" regime. A better comparison is at $V_5 = 600$ km s$^{-1}$. NKP94 used this to (prematurely) claim CfA1 groups had significantly cooler cores than any simulations.

Figure 18. The same as Figure 17, but using the breakup catalogs and making the N(z) correction to the median $v_{\rm gr}$'s. The "cooler cores" conclusion of NKP94 is now seen to be an artifact of overmerging and N(z) bias. When corrected, CHDM is in good agreement with observations at both fiducial $V_5$'s, while CDM remains too high.

Figure 19. The same as Figure 18, but now using the CfA-sky catalogs. The error bars are larger, but otherwise the figure is virtually identical to that for the full 10.384 sr catalogs. At low and moderate $V_5$, $f_{\rm gr}$ differs very little between these samples.

Figure 20. Percolation properties of groups vs. varying $V_5$ in the fully corrected case for the CfA-sky catalogs. CDM produces too few groups at low $V_5$, while CHDM$_2$ produces too many at intermediate and high $V_5$. CHDM$_1$ percolates too easily, as groups merge faster than they are created above $V_5 = 450$ km s$^{-1}$. A curve intermediate between CHDM$_1$ and CHDM$_2$ would likely give the best fit, though discrimination between models is poor in this plane.

Figure 21. Percolation properties of groups vs. varying $D_n$ for all simulations, using the CfA-sky catalogs. As in Figure 20, CHDM$_1$ percolates too easily. CHDM$_2$ percolates correctly, but produces too many groups at all $D_n$. A CHDM model with large scale power intermediate between these, and lower $\Omega_\nu$ would appear to fit better. Both CDM models fit well on this measure.

Figure 22. The ratio of the simulation mean harmonic radius $r_h$ to that for the CfA1 at the same links, vs. $D_n$. All simulations are significantly too high. This may be due to residual resolution limitations, or to simulation halos having spatial distributions not sufficiently galaxy-like.

This figure "fig3-1.png" is available in "png" format from:

http://arxiv.org/ps/astro-ph/9410095v1

This figure "fig1-3.png" is available in "png" format from:

http://arxiv.org/ps/astro-ph/9410095v1